\documentclass{article}
\usepackage{cite,bm,graphicx}

\title{Stochastic analysis of surface roughness}

\author{M.~Waechter$^{1,2}$\thanks{%
     E-mail: \texttt{matthias.waechter@uni-oldenburg.de}}
     \and F.~Riess$^{1}$
     \and H.~Kantz$^{2}$
     \and  J.~Peinke$^{1}$\thanks{%
     E-mail: \texttt{peinke@uni-oldenburg.de}}
}
\date{\small
   $^{1}$ Carl von Ossietzky University, Physics Department -
            D-26111 Oldenburg, Germany\\
   $^{2}$ Max Planck Institute for Physics of Complex Systems -
            D-01187 Dresden, Germany\\[1ex]
            \normalsize 9.9.2003
}
\newcommand{\un}[1]{\ensuremath{\,\mathrm{#1}}}
\newlength{\breite}

\begin{document}
\maketitle

\begin{abstract}
  
  For the characterization of surface height profiles we present a new
  stochastic approach which is based on the theory of Markov processes.  With
  this analysis we achieve a characterization of the complexity of the surface
  roughness by means of a Fokker-Planck or Langevin equation, providing the
  complete stochastic information of multiscale joint probabilities.  The
  method was applied to different road surface profiles which were measured
  with high resolution.  Evidence of Markov properties is shown. Estimations
  for the parameters of the Fokker-Planck equation are based on pure,
  parameter free data analysis.

\end{abstract}

\section{Introduction}\label{sec:intro}

The complexity of rough surfaces is subject of a large variety of
investigations in different fields of science 
\cite{Sayles1978, Vicsek1992-e, Barabasi1995, Davies1999}.
Physical and chemical properties of surfaces and interfaces are to a
significant degree determined by their topographic structure.
As one example, the influence of surface roughness on boundary layer flows is
discussed in turbulence research (cf.\ \cite{Keirsbulck2002, Smalley2002}) and
in the atmospheric (cf.\ \cite{Stull1988, Roth2000}) and oceanographic
sciences (cf.\ \cite{Vanneste2003}).
A comprehensive characterization of the topography is of vital interest 
for deposited, polished or otherwise processed surfaces. Therefore, to give
a second example, current roughness analysis methods find practical application
for the characterization of polished surfaces \cite{Dharmadhikari1999,
  Saitou2001, Sydow2003}. 

In the context of industrial and engineerig applications common roughness
measures are based on standardized procedures as the maximum height difference
$R_z$ and the mean value of absolute heights $R_a$ \cite{ISO4287}. These
rather simple measures clearly cannot completely characterize the complexity
of roughness. This is also confirmed by the existence of a large amount of
additional measures, each describing a very special feature of a surface.

In the physical and other sciences a common local measure of roughness is the
rms surface width $w_r(x)=\langle (h(\tilde{x})-\bar{h})^2 \rangle^{1/2}_r$,
where $h(\tilde{x})$ is the measured height at point $\tilde{x}$,
$\langle\,\cdot\,\rangle_r$ denotes the average over an interval of length $r$
around the selected point $x$, and $\bar{h}$ the mean value of $h(\tilde{x})$
in that interval. Advantages of $w_r(x)$ are a scale dependent definition as
well as clear physical and stochastical meanings.
Among the techniques used to characterize scale dependent surface roughness
probably the most prominent ones are the concepts of self-affinity and
multi-affinity, where the multi-affine $f(\alpha)$ spectrum has been regarded
as the most complete characterization of a surface \cite{Feder1988,
  Family1991, Vicsek1992-e, Barabasi1995}.
From a stochastic point of view we want to point out two important properties 
of multiaffinity: %
(a) the ensemble average $\langle w_r \rangle$ must obey a scaling law
$\langle w_r^\alpha \rangle \sim r^{\xi_\alpha}$, and %
(b) the statistics of $w_r(x)$ are investigated on distinct length scales,
thus possible correlations between $w_r(x)$ and $w_{r'}(x)$
on different scales $r,r'$ are not examined.

The method we are proposing in this contribution is based on stochastic
processes which should grasp the scale dependency of surface roughness in a
most general way. No scaling feature is explicitly required, and especially the
correlations between different scales $r$ and $r'$ are investigated.
To this end we present a systematic procedure how the explicit form of a
stochastic process for the $r$-evolution of a roughness measure similar to
$w_r(x)$ can be extracted directly from the measured surface topography.
This stochastic approach has turned out to be a promising tool also for other
systems with scale dependent complexity like turbulence \cite{Renner2001,
  Friedrich1998a} and financial data \cite{Friedrich2000a}.
In this letter we focus on the complexity of rough surfaces.  As a specialized
example we have picked out the applied problem of characterizing road
surfaces.  These are an essential component of current transportation and thus
represent a class of non-idealized and widely used surfaces.
We claim that the applicability of our method to this class of surfaces
indicates its general relevance for improved surface characterization.
%

A collection of road surface data was measured which served previously as an
empirical basis for the prediction of vibrational stress on bicycle riders
\cite{Waechter2002a}. It is common to describe the quality of a road surface
by a power law fit to the power spectrum of the height profile
\cite{Braun1969-e, Dodds1973, ISO8608}.  This method is not appropriate
especially for wavelengths below 0.3\,m and for non-Gaussian height
distributions. 
Furthermore, the power spectrum characterizes only the $r$-dependence of one
moment of the two-point correlations \cite{SecondMom}.
While some improvements to this method have been proposed \cite{Bruscella1999}
the characterization of road surfaces still remains incomplete
\cite{Kempkens1999-e,Ueckermann1999-e}.  
For the improvement of road surface characterization multifractality and
multiaffinity seem not to be appropriate tools because scaling is no
constant feature of road surfaces.  
%

In the remainder of this letter we first introduce our method, the
determination of a Fokker-Planck equation for the evolution of conditional
probability density functions (pdf) directly from experimental data. Next, we
present a typical data set, show evidence of its Markov properties, and
estimate the coefficients of the Fokker-Planck equation. At last, we evaluate
the precision of the estimated coefficients by numerical reconstruction of
conditional and unconditional pdfs.

\section{Method}\label{sec:method}

It is one common procedure to characterize the complexity of a rough
surface by the statistics of the \emph{height increment}\cite{CenteredInc} 
\begin{equation}
  \label{eq:h_r}
  h_r(x) := h(x+r/2) - h(x-r/2)
\end{equation}
depending on the length scale $r$, as marked in fig.~\ref{fig:data}.
Other scale dependent roughness measures can, for example, be found in
\cite{Family1991,Barabasi1995}. Here we use the height increment $h_r$ because
it is also directly linked to vehicle vibrations induced by the road surface if
$r$ is the wheelbase. Another argument for the use of $h_r$ is that its moments
are connected with spatial correlation functions, but it should be pointed out
that our method presented in the following could be easily generalized to any
scale dependent measure, \emph{e.g.} the above-mentioned $w_r(x)$. As a new
ansatz, $h_r$ will be regarded here as a \emph{stochastic variable in $r$}.
Without loss of generality we consider the process as being directed from
larger to smaller scales.  Our interest is the investigation how surface
roughness is linked between different length scales.

Complete information about the stochastic process would be available by the
knowledge of all possible $n$-point, or more precisely $n$-scale,
probability density functions (pdf) %
$p(h_1, r_1; h_2, r_2; \ldots ; h_n, r_n)$ %
describing the probability of finding simultaneously the increments $h_1$ on
the scale $r_1$, $h_2$ on the scale $r_2$, and so forth up to $h_n$ on the
scale $r_n$. Here we use the notation $h_i(x)=h_{r_i}(x)$, see (\ref{eq:h_r}).
Without loss of generality we take $r_1<r_2<\ldots<r_n$. As a first question
one has to ask for a suitable simplification. In any case the $n$-scale joint
pdf can be expressed by multiconditional pdf
\begin{eqnarray} \label{eq:condpdf}
   p(h_1,r_1;\ldots;h_n,r_n) &=& p(h_1,r_1 | h_2,r_2;\ldots;h_n,r_n)
                             \cdot p(h_2,r_2|h_3,r_3;\ldots;h_n,r_n)\nonumber\\
     &&\cdot\ldots\cdot p(h_{n-1},r_{n-1}|h_n,r_n)\cdot p(h_n,r_n)
\end{eqnarray}
where $p(h_i,r_i| h_j,r_j)$ denotes a \emph{conditional probability}, which is
defined as the probability of finding the increment $h_i$ on the scale $r_i$
under the condition that simultaneously, \emph{i.e.}\ at the same location
$x$, on a larger scale $r_j$ the value $h_j$ was found.  An important
simplification arises if
\begin{equation}
   \label{eq:markov_straight}
   p(h_i, r_i | h_{i+1}, r_{i+1}; \ldots; h_n, r_n) =
   p(h_i, r_i | h_{i+1}, r_{i+1})\;.
\end{equation}
This property is the defining feature of a Markov process evolving from
$r_{i+1}$ to $r_i$. Thus for a Markov process the $n$-scale joint pdf
factorize into $n$ conditional pdf
\begin{equation}
   \label{eq:markov1}
       p(h_1,r_1;\ldots;h_n,r_n) =
       p(h_1,r_1\, | h_2,r_2)\cdot \ldots
       \cdot p(h_{n-1},r_{n-1}\, | h_n,r_n)
       \cdot p(h_n,r_n) \;.
\end{equation}
This Markov property implies that the $r$-dependence of $h_r$ can be regarded
as a stochastic process evolving in $r$, driven by deterministic and random
forces. If additionally the included noise is Gaussian distributed, the
process can be described by a Fokker-Planck equation \cite{Risken1984}.  For
our height profiles it takes the form
\begin{equation}
   \label{eq:FPE1}
     -r\,\frac{\partial}{\partial r}\;p(h_r,r|h_0,r_0)\,=
     \left\{ -\frac{\partial}{\partial h_r} D^{(1)}(h_r,r)
       + \frac{\partial^2}{\partial h_r^2} D^{(2)}(h_r,r)
     \right\} \, p(h_r,r|h_0,r_0) \;.
\end{equation}
The Fokker-Planck equation then describes the evolution of the conditional
probability density function from larger to smaller length scales and 
thus also the complete $n$-scale statistics.
The minus sign on the left side of eq.~(\ref{eq:FPE1}) expresses this
direction of the process, furthermore the factor $r$ corresponds to a
logarithmic variable $\rho=\ln r$ which leads to simplified results in the
case of scaling behaviour \cite{FPE}.

The term $D^{(1)}(h_r,r)$ is commonly denoted as drift term, describing the
deterministic part of the process, and $D^{(2)}(h_r,r)$ as diffusion term being
the variance of a Gaussian, $\delta$-correlated noise.  Focus of our analysis
is to show evidence of the above mentioned Markov property and to derive the
drift and diffusion coefficients $D^{(1)}$ and $D^{(2)}$ in eq.~(\ref{eq:FPE1})
from experimental data.

\section{Data}\label{sec:data}

\setlength{\breite}{0.48\linewidth}
\begin{figure}[htbp]
  \parbox{\breite}{%
    \includegraphics[width=\linewidth]{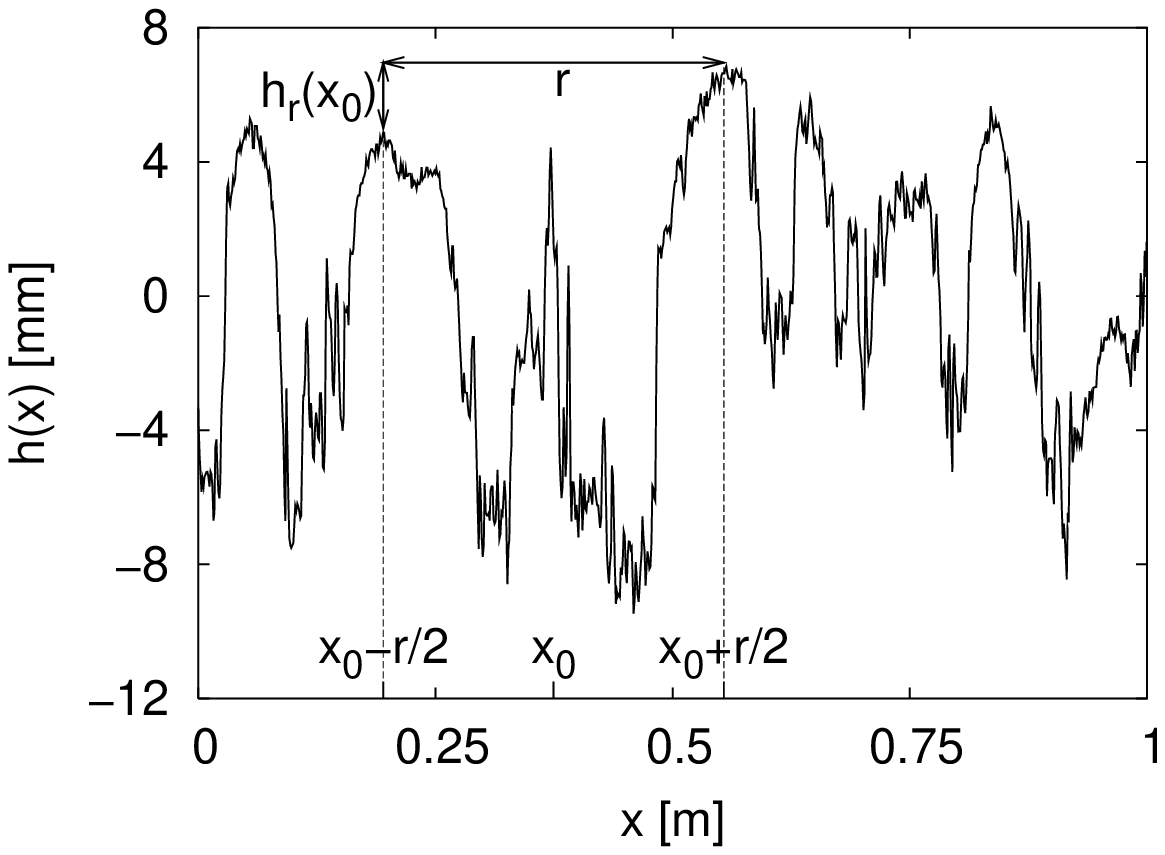}
    \caption{Cut-out from a height profile of an irregular cobblestone road.
      Additionally the construction of the height increment $h_r(x_0) =
      h(x_0+r/2) - h(x_0-r/2)$ is illustrated.}
    \label{fig:data}
    }
  \hspace*{\fill}
  \parbox{\breite}{%
    \includegraphics[width=\linewidth]{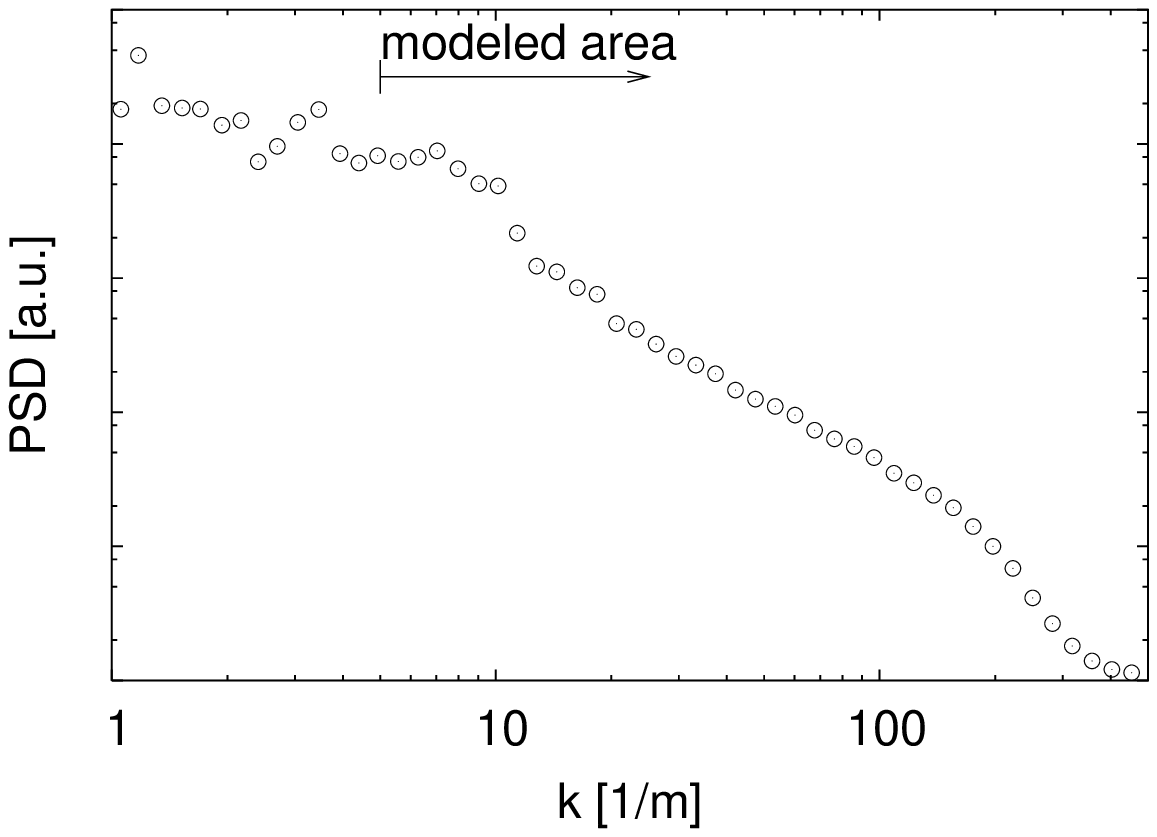}
    \caption{Power spectral density of the data set. The arrow indicates
      beginning and direction of the range where drift and diffusion
      coefficient could be obtained and verified.}
    \label{fig:psd}
    }
\end{figure}
Height profiles were measured from numerous road and cycle track surfaces
typical for West German bicycle traffic. Road sections were selected 
in such a way that stationarity is given.
Profile length is typically $20\un{m}$ or 19\,000 samples, respectively.
Longitudinal resolution was $1.04\un{mm}$ and vertical error smaller than
$0.5\un{mm}$. 
As an example we present results from a data set of an irregular cobblestone
road consisting of ten profiles with a total of about 190\,000 samples
\cite{Strassen-Daten}. 
Figure~\ref{fig:data} shows a short section of the data.
In fig.~\ref{fig:psd} the power spectral density of the height profiles is
plotted against the wavenumber. Scaling behaviour is not found at the
beginning of the analysed region of length scales, while for smaller scales 
($20 < k/\un{m^{-1}} < 100$) it appears to be present.
In the following, the height increments are normalized by
\mbox{$\sigma_\infty:=\lim_{r\rightarrow\infty}\langle h_r^2\rangle^{1/2}$},
with $\sigma_\infty=6.3\un{mm}$ for the given data set.

\section{Markov Properties}\label{sec:markov}

For a Markov process the defining feature is that the $n$-point conditional
probability distributions are equal to the single conditional probabilities,
according to eq.~(\ref{eq:markov_straight}). With the given amount of
data points the verification of this condition is only possible for $n=3$ and
for $r_1<r_2<r_3<300\un{mm}$.

\begin{figure}[htbp]
  \begin{center}
  \includegraphics[width=0.8\linewidth]{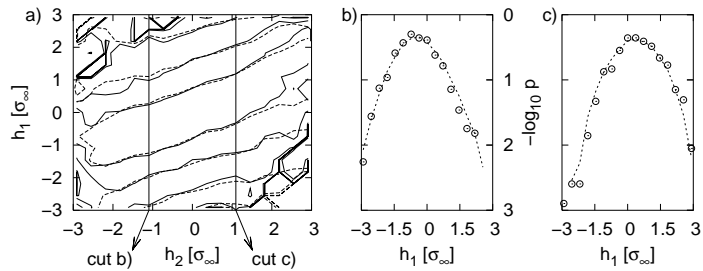}
  \caption{a) Contour plot of single and double conditional probabilities
    $p(h_1,r_1 | h_2,r_2)$ (dashed lines) and $p(h_1,r_1 | h_2,r_2;
    h_3\!\!=\!\!0,r_3)$ (solid lines) for $r_1=126; r_2=188; r_3=251\un{mm}$.
    Contour levels differ by a factor of 10, with an additional level at
    $p=0.3$.
    b), c) Two one-dimensional cuts at $h_2 \approx \pm \sigma_\infty$ with
    $p(h_1,r_1 | h_2,r_2)$ as dashed lines and $p(h_1,r_1 | h_2,r_2; h_3,r_3)$
    as circles.}
  \label{fig:markov}
  \end{center}
\end{figure}

Figure~\ref{fig:markov} shows both sides of eq.~(\ref{eq:markov_straight})
with $n=3$ in a contour plot and two cuts at $h_2\approx\pm\sigma_\infty$. The
value of $h_3$ was chosen to be $h_3=0$. We take this rather good
correspondence as a strong hint for a Markov process. Markov properties were
found for scale distances from about $17\un{mm}$ upwards. Note that the main
axis of the distribution is tilted, indicating that $p(h_1,r_1 | h_2,r_2) \neq
p(h_1,r_1)$ and thus height increments on different scales are not
independent.

\section{Estimation of Drift and Diffusion Coefficients}

In order to obtain the drift ($D^{(1)}$) and diffusion coefficient ($D^{(2)}$)
for eq.~(\ref{eq:FPE1}) we proceed in a well defined way like it was already
expressed by Kolmogorov \cite{Kolmogorov1931}, see also
\cite{Risken1984,Renner2001}. First, the conditional moments
$M^{(k)}(h_r,r,\Delta_r)$ for finite step sizes $\Delta r$ are directly
estimated from the data via moments of the conditional probabilities
\begin{equation}
   \label{eq:Mk}
   M^{(k)}(h_r,r,\Delta r) =
     \frac{r}{k!\Delta r}
     \int_{\scriptscriptstyle-\infty}^{\scriptscriptstyle+\infty}
     (\tilde{h}-h_r)^k \; p(\tilde{h},r-\Delta r | h_r,r) \;  d\tilde{h} \;.
\end{equation}
Second, the coefficients $D^{(k)}(h_r,r)$ are obtained from the limit of
$M^{(k)}(h_r,r,\Delta_r)$ when $\Delta r$ approaches zero
\begin{equation}
  \label{eq:Dk}
    D^{(k)}(h_r,r) = \lim_{\Delta r \rightarrow 0} M^{(k)}(h_r,r,\Delta r) \;.
\end{equation}

\begin{figure}[htbp]
  \hspace*{\fill}
  \includegraphics[width=0.4\linewidth]{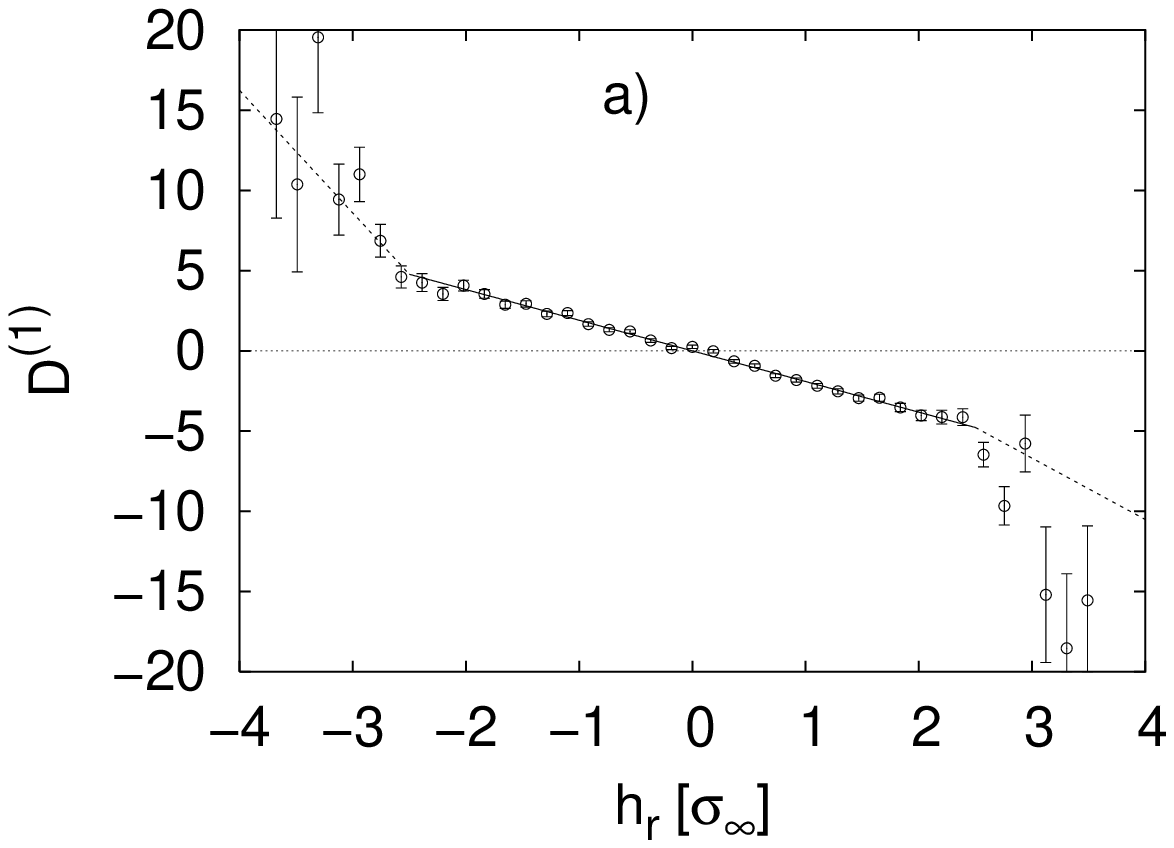}
  \hspace*{\fill}
  \includegraphics[width=0.4\linewidth]{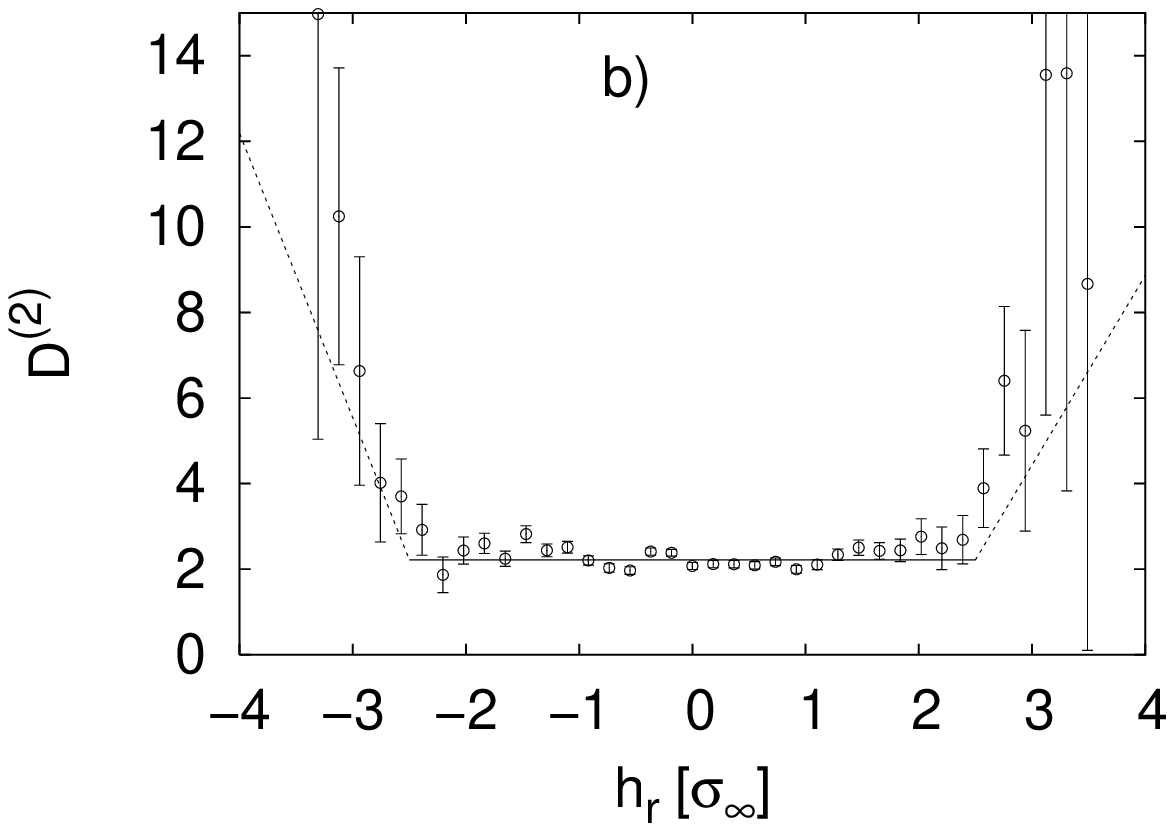}
  \hspace*{\fill}
  \caption[Estimated drift coefficient $D^{(1)}(h_r, r)$]%
  {Estimated coefficients of the Fokker-Planck equation for $r=188$\,mm.
    Parameterizations are shown as solid and broken lines.  a) Drift
    coefficient $D^{(1)}(h_r, r)$. b) Diffusion coefficient $D^{(2)}(h_r, r)$.}
  \label{fig:D1}\label{fig:D2}
\end{figure}
Figure~\ref{fig:D1} shows estimations of the drift coefficient $D^{(1)}$ and
the diffusion coefficient $D^{(2)}$ for $r=188$\,mm. The error bars are
estimated from the errors of $M^{(1)}$ and $M^{(2)}$ via the number of events
contributing to each value. The limit $\Delta r \rightarrow 0$ was performed in
both cases by a linear fit to $M^{(k)}(h_r,r,\Delta r)$ in the range
$17\un{mm}\leq\Delta r\leq29\un{mm}$. Both coefficients were parameterized as
piecewise linear functions where the behaviour within
$|h_r|\leq2.5\sigma_\infty$ could be derived directly from the above
estimations. Outside this range increasing errors make a precise estimate more
difficult. Parameterizations were chosen here to additionally obtain good
results in the verification procedures (see below). Figure~\ref{fig:D1} shows
that the resulting parameterizations are in good agreement with the
estimations.
It is easy to verify that with linear $D^{(1)}$ and constant $D^{(2)}$ the
Fokker-Planck equation~(\ref{eq:FPE1}) describes a Gaussian process, while with
a parabolic $D^{(2)}$ the distributions become non-Gaussian, also called
intermittent or heavy tailed.

In our case, here, $D^{(1)}(h_r,r)$ is characterized by the slope $-\gamma(r)$
within $|h_r|\leq2.5\sigma_\infty$, $-4\gamma(r)$ for $h_r<-2.5\sigma_\infty$,
and $-2\gamma(r)$ for $h_r>2.5\sigma_\infty$ (compare fig.~\ref{fig:D1}). The
dependence of $\gamma$ on $r$ is nontrivial with the value ranging from 0.82
for $r=83$\,mm to 1.9 for $r=188$\,mm.
$D^{(2)}(h_r,r)$ was found to have a value $\beta(r)$ independent of $h_r$
within $|h_r|\leq2.5\sigma_\infty$. For $h_r<-2.5\sigma_\infty$ $D^{(2)}$ was
parameterized to be linear with slope $-4\beta(r)$, for $h_r>2.5\sigma_\infty$
with slope $3\beta(r)$. The dependence of $\beta$ on $r$ can be approximated
by $\beta(r)=0.0117\, r/\mbox{mm}$.

\section{Verification of Coefficients}

\begin{figure}[htbp]
  \parbox{\breite}{%
    \begin{center}
      \includegraphics[width=0.7\linewidth]{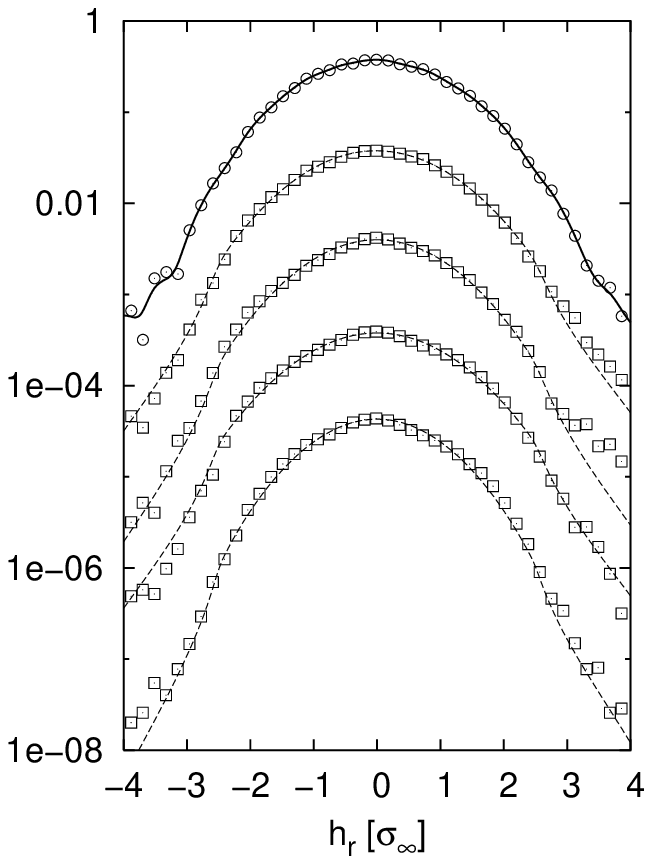}
      \caption{Numerical solution of the integrated form of Fokker-Planck equation
        (\ref{eq:FPE1}) compared to empirical pdf (symbols) at different scales 
        $r$. Solid line: empirical pdf parameterized at $r=188\un{mm}$, dashed
        lines: reconstructed pdf. Scales $r$ are 188, 158, 112, 79, 46\un{mm} from
        top to bottom. Pdf are shifted in vertical direction for clarity of
        presentation.}
      \label{fig:veri1}
    \end{center}
    }
  \hspace*{\fill}
  \parbox{\breite}{%
    \begin{center}
    \includegraphics[width=0.8\linewidth]{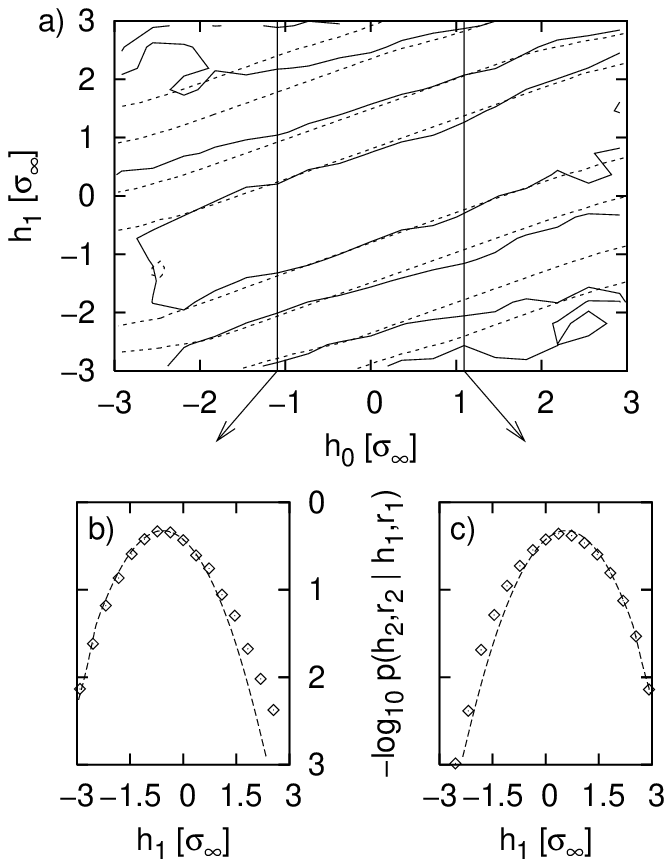}
    \caption{Direct numerical solution of Fokker-Planck equation (\ref{eq:FPE1})
      compared to the empirical pdf at scales $r_0=188$, $r_1=131$\,mm.  a)
      Contour plot of empirical (solid lines) and reconstructed pdf (dashed
      lines). Contour levels are as in fig.~\ref{fig:markov}.
      b,c) Cuts at $h_0\approx\pm\sigma_\infty$. Empirical pdf are plotted 
      as symbols.}
    \label{fig:veri2}
    \end{center}
    }
\end{figure}

Next, we want to evaluate the precision of our result. Therefore we return to
eq.~(\ref{eq:FPE1}). Knowing $D^{(1)}$ and $D^{(2)}$ it should be possible
to calculate the pdf of $h_r$ with the corresponding Fokker-Planck equation.
Equation~(\ref{eq:FPE1}) can be integrated over $h_0$ and is then valid also
for the unconditional pdf. 
Now the empirical pdf at $r_0=188\un{mm}$ is parameterized (see
fig.~\ref{fig:veri1}) and used as initial condition for a numerical solution
of the integrated form of eq.~(\ref{eq:FPE1}). For several values of $r$ the
reconstructed pdf is compared to the respective empirical pdf, as shown in
fig.~\ref{fig:veri1}. Please note that the interchange of steeper and flatter
regions in the reconstructed pdf is achieved by the piecewise linear
parameterization of $D^{(1)}$ and $D^{(2)}$.

A second verification is the reconstruction of conditional pdf by direct
numerical solution of the Fokker-Planck equation (\ref{eq:FPE1}). An example
for the scales $r_0=188$\,mm and $r_1=131$\,mm is shown in
fig.~\ref{fig:veri2}. 
Reconstructing the conditional pdf this way is much more sensitive to
deviations in $D^{(1)}$ and $D^{(2)}$. 
This becomes evident by the fact that the conditional pdf (and not the
unconditional pdf of fig.~\ref{fig:veri1}) determine $D^{(1)}$ and $D^{(2)}$
(see (\ref{eq:Mk}) and (\ref{eq:Dk})). Here also the difference to
the multiscaling analysis becomes clear, which analyses higher moments
$\langle h_r^q\rangle = \int h_r^q\cdot p(h_r)\,\mathrm{d} h_r$ of $h_r$, and does
not depend on conditional pdf. It is easy to show that there are many
different stochastic processes which lead to the same single scale pdf
$p(h_r)$.

\section{Conclusions}

The height increment $h_r$ of surface height profiles as a stochastic variable
in $r$ can be correctly described by a Fokker-Planck equation with drift and
diffusion coefficients derived directly from measured data.  The results of
the presented example support the hypothesis that the noise term in the
evolution of the stochastic variable $h_r$ in $r$ is sufficiently well
described by a Gaussian, $\delta$-correlated random process.

As the Fokker-Planck equation describes the evolution of $p(h_r,r\,|\,h_0,
r_0)$ and $p(h_r,r)$ with $r$, it covers also the behaviour of the moments
$\langle h_r^n\rangle$ including any possible scaling behaviour. From the
integrated form of eq.~(\ref{eq:FPE1}) an equation for the moments can be
obtained by multiplying with $h_r^n$ and integrating over $h_r$.
For $D^{(1)}$ being purely linear in $h_r$ 
and $D^{(2)}$ purely quadratic, 
multifractal scaling 
is obtained. 
%
We note again that, compared to scaling features, the knowledge of the
Fokker-Planck equation provides more information on the complexity of the
surface roughness in the sense of multi-scale joint probability density
functions, eq.~(\ref{eq:condpdf}), which correspond to multipoint statistics.
While to this end we do not seem to find universal laws concerning rough
structures, we do achieve a comprehensive characterization of a specific
surface, showing the strength and generality of this method.

At last we want to point out that the Fokker-Planck equation (\ref{eq:FPE1})
corresponds to an equivalent Langevin equation \cite{Risken1984}.
The use of this Langevin equation in the scale variable should
open the possibility to directly simulate surface profiles with given
stochastic properties for different applications. 

Financial support by the Volkswagen Foundation is kindly acknowledged.

\bibliography{mw,obf-letter}

\begin{thebibliography}{10}
\bibitem{SecondMom}
The Fourier transform of the power spectrum $\langle h(x+r/2)h(x-r/2)\rangle$
  can easily be obtained from the second moment $\langle
  (h(x+r/2)-h(x-r/2))^2\rangle$ of $h_r(x)$ defined below.
\bibitem{CenteredInc}
Please note that there have been different definitions of increments,
  especially the left-justified increment $h_r(x)=h(x+r)-h(x)$.
\bibitem{Strassen-Daten}
The data have been high-pass filtered with a cutoff wavelength of 1~m. Due to
  the limited profile length effects on longer scales cannot be analysed in a
  stochastical sense. 
  However, this limitation to lengthscales below typical vehicle wheelbases
  does not mean a principal drawback of the proposed analysis.
\end{thebibliography}
\bibliographystyle{unsrt}

\end{document}